\documentclass[longauth]{aa}
\usepackage{graphicx}
\usepackage{txfonts}
\usepackage{xcolor}
\usepackage{hyperref}
\hypersetup{colorlinks, linkcolor=blue, citecolor=blue, urlcolor=blue}
\newcommand{\pivec}{\mbox{\boldmath $\pi$}}
\newcommand{\muvec}{\mbox{\boldmath $\mu$}}

\newcommand{\te}{t_{\rm E}}
\newcommand{\thetae}{\theta_{\rm E}}

\newcommand{\pie}{\pi_{\rm E}}
\newcommand{\pien}{\pi_{{\rm E},N}}
\newcommand{\piee}{\pi_{{\rm E},E}}
\newcommand{\dl}{D_{\rm L}}

\definecolor{brown}{rgb}{0.59, 0.29, 0.0}
\definecolor{darkgreen}{rgb}{0.0, 0.42, 0.24}
\definecolor{darkblue}{rgb}{0.01, 0.31, 0.59}
\definecolor{darkblue}{rgb}{0.0, 0.25, 0.42}
\definecolor{blue}{rgb}{0.0,0.0,1.0}
\definecolor{green}{rgb}{0.0,1.0,0.0}

\begin{document}

%\title{Multi-planet microlensing system OGLE-2019-BLG-0468L composed of two giant planets orbiting around a G-type star}
%\title{Brown dwarf companions in microlensing binaries detected during the 2018--2020 season KMTNet data}
\title{Probable brown dwarf companions detected in binary microlensing events during the 2018-2020 seasons of the KMTNet survey}
\titlerunning{Brown dwarf companions in microlensing binaries}

\author{
% leading author -----------------------------
     Cheongho Han\inst{01}
\and Youn Kil Jung\inst{02}
\and Doeon Kim\inst{01}
\and Andrew Gould\inst{03,04}
\and Valerio Bozza\inst{05,06}
%\and Andrzej Udalski\inst{08}
\and Ian~A.~Bond\inst{07}
\\
(Leading authors)\\
% KMTNet ---------------------------
     Sun-Ju Chung\inst{02}
\and Michael D. Albrow\inst{08}
\and Kyu-Ha Hwang\inst{02}
\and Yoon-Hyun Ryu\inst{02}
\and In-Gu Shin\inst{09}
\and Yossi Shvartzvald\inst{10}
\and Hongjing Yang\inst{11}
\and Weicheng Zang\inst{09,11}
\and Sang-Mok Cha\inst{02,12}
\and Dong-Jin Kim\inst{02}
\and Hyoun-Woo Kim\inst{02}
\and Seung-Lee Kim\inst{02}
\and Chung-Uk Lee\inst{02}
\and Dong-Joo Lee\inst{02}
\and Jennifer~C.~Yee\inst{09}
\and Yongseok Lee\inst{02,13}
\and Byeong-Gon Park\inst{02,13}
\and Richard W. Pogge\inst{04}
\\
(The KMTNet collaboration)\\
% OGLE ---------------------------
%     Przemek Mr{\'o}z\inst{08}
%\and Micha{\l} K. Szyma{\'n}ski\inst{08}
%\and Jan Skowron\inst{08}
%\and Radek Poleski\inst{08}
%\and Igor Soszy{\'n}ski\inst{08}
%\and Pawe{\l} Pietrukowicz\inst{08}
%\and Szymon Koz{\l}owski\inst{08}
%\and Krzysztof Ulaczyk\inst{16}
%\and Krzysztof A. Rybicki\inst{08,12}
%\and Patryk Iwanek\inst{08}
%\and Marcin Wrona\inst{08}
%\\
%(The OGLE Collaboration)\\
%% MOA ---------------------------
     Fumio Abe\inst{14}
\and Richard Barry\inst{15}
\and David P. Bennett\inst{15,16}
\and Aparna Bhattacharya\inst{15,16}
\and Hirosame Fujii\inst{14}
\and Akihiko~Fukui\inst{17,18}
\and Yuki Hirao\inst{19}
\and Stela Ishitani Silva\inst{16,21}
\and Rintaro Kirikawa\inst{19}
\and Iona Kondo\inst{19}
\and Naoki Koshimoto\inst{22}
\and Yutaka Matsubara\inst{14}
\and Sho~Matsumoto\inst{19}
\and Shota Miyazaki\inst{19}
\and Yasushi Muraki\inst{14}
\and Arisa Okamura\inst{19}
\and Greg Olmschenk\inst{16}
\and Cl{\'e}ment Ranc\inst{23}
\and Nicholas J. Rattenbury\inst{24}
\and Yuki Satoh\inst{19}
\and Takahiro Sumi\inst{19}
\and Daisuke Suzuki\inst{19}
\and Taiga Toda\inst{19}
\and Paul~J.~Tristram\inst{25}
\and Aikaterini Vandorou\inst{15,16}
\and Hibiki Yama\inst{19}
\and Yoshitaka Itow\inst{14}
\\
(The MOA Collaboration)\\
}

\institute{
      Department of Physics, Chungbuk National University, Cheongju 28644, Republic of Korea,                                                            % (01)
\and  Korea Astronomy and Space Science Institute, Daejon 34055, Republic of Korea                                                                       % (02)
\and  Max-Planck-Institute for Astronomy, K\"{o}nigstuhl 17, 69117 Heidelberg, Germany                                                                   % (03)
\and  Department of Astronomy, Ohio State University, 140 W. 18th Ave., Columbus, OH 43210, USA                                                          % (04)
\and  Dipartimento di Fisica "E. R. Caianiello", Universit\'a di Salerno, Via Giovanni Paolo II, 84084 Fisciano (SA), Italy                              % (05)
\and  Istituto Nazionale di Fisica Nucleare, Sezione di Napoli, Via Cintia, 80126 Napoli, Italy                                                          % (06)
\and  Institute of Natural and Mathematical Science, Massey University, Auckland 0745, New Zealand                                                       % (07)
\and  University of Canterbury, Department of Physics and Astronomy, Private Bag 4800, Christchurch 8020, New Zealand                                    % (08)
\and  Center for Astrophysics $|$ Harvard \& Smithsonian, 60 Garden St., Cambridge, MA 02138, USA                                                        % (09)
\and  Department of Particle Physics and Astrophysics, Weizmann Institute of Science, Rehovot 76100, Israel                                              % (10)
\and  Department of Astronomy, Tsinghua University, Beijing 100084, China                                                                                % (11)
\and  School of Space Research, Kyung Hee University, Yongin, Kyeonggi 17104, Republic of Korea                                                          % (12)
\and  Korea University of Science and Technology, Korea, (UST), 217 Gajeong-ro, Yuseong-gu, Daejeon, 34113, Republic of Korea                            % (13)
\and  Institute for Space-Earth Environmental Research, Nagoya University, Nagoya 464-8601, Japan                                                        % (14)
\and  Code 667, NASA Goddard Space Flight Center, Greenbelt, MD 20771, USA                                                                               % (15)
\and  Department of Astronomy, University of Maryland, College Park, MD 20742, USA                                                                       % (16)
\and  Department of Earth and Planetary Science, Graduate School of Science, The University of Tokyo, 7-3-1 Hongo, Bunkyo-ku, Tokyo 113-0033, Japan      % (17)
\and  Instituto de Astrof{\'i}sica de Canarias, V{\'i}a L{\'a}ctea s/n, E-38205 La Laguna, Tenerife, Spain                                               % (18)
\and  Department of Earth and Space Science, Graduate School of Science, Osaka University, Toyonaka, Osaka 560-0043, Japan                               % (19)
\and  Department of Physics, The Catholic University of America, Washington, DC 20064, USA                                                               % (20)
\and  Department of Physics, The Catholic University of America, Washington, DC 20064, USA                                                               % (21)
\and  Department of Astronomy, Graduate School of Science, The University of Tokyo, 7-3-1 Hongo, Bunkyo-ku, Tokyo 113-0033, Japan                        % (22)
\and  Zentrum f{\"u}r Astronomie der Universit{\"a}t Heidelberg, Astronomisches Rechen-Institut, M{\"o}nchhofstr.\ 12-14, 69120 Heidelberg, Germany      % (23)
\and  Department of Physics, University of Auckland, Private Bag 92019, Auckland, New Zealand                                                            % (24)
\and  University of Canterbury Mt. John Observatory, P.O. Box 56, Lake Tekapo 8770, New Zealand                                                          % (25)
}
%\date{Received ; accepted}
%

% \abstract{}{}{}{}{} 
% 5 {} token are mandatory
\abstract
{}
{
We inspect the microlensing data of the KMTNet survey collected during the 2018--2020
seasons in order to find lensing events produced by binaries with brown-dwarf companions.
}
{
In order to pick out binary-lens events with candidate BD lens companions, we conduct
systematic analyses of all anomalous lensing events observed during the seasons. By applying the
selection criterion with mass ratio between the lens components of $0.03\lesssim q\lesssim 0.1$, 
we identify  four binary-lens events with candidate BD companions, including KMT-2018-BLG-0321,
KMT-2018-BLG-0885, KMT-2019-BLG-0297, and KMT-2019-BLG-0335.  For the individual events, we present 
the interpretations of the lens systems and measure the observables that can constrain the physical 
lens parameters.
}
{
The masses of the lens companions estimated from the Bayesian analyses based on the measured
observables indicate that the probabilities for the lens companions to be in the brown-dwarf 
mass regime are high: 59\%, 68\%, 66\%, and 66\% for the four events respectively.
}
{}


\maketitle

\section{Introduction}\label{sec:one}

Due to the trait of occurring by the mass of a lensing object regardless of its luminosity, 
microlensing provides an important tool to detect very faint and even dark astronomical objects.  
With this trait, microlensing has been applied to search for extrasolar planets, and about 30 
planets are annually being detected \citep{Gould2022b, Jung2022, Gould2022a} from the combined 
observations by survey, for example, the KMTNet \citep{Kim2016}, MOA \citep{Bond2001}, and OGLE 
\citep{Udalski2015} experiments, and followup groups, for example, the ROME/REA survey 
\citep{Tsapras2019}.

Brown dwarfs (BDs) are another population of faint astronomical objects to which microlensing 
is sensitive.  Microlensing BDs can be detected through a single-lens event channel, in which 
a single BD object produces a lensing event with a short time scale, for example, \citet{Han2020}. 
Considering that BDs may have formed via a similar mechanism to that of stars, BDs can be as abundant 
as its stellar siblings, and, in this case, a significant fraction of short time-scale lensing events 
being detected by the surveys may be produced by BDs.  Observationally, both radial-velocity 
\citep{Grether2006} and microlensing \citep{Shvartzvald2016} studies indicate a deficit of BDs 
companions compared to both stars and planets, suggesting that microlensing may allows us to study 
the stars, BDs and planet formation mechanism.  However, confirming the BD lens nature of a short 
time-scale event by measuring the mass of the lens is difficult because the event time scale depends 
not only on the lens mass but also on the distance to the lens and the relative lens-source proper 
motion. The lens mass can be determined by simultaneously measuring the extra lensing observables 
of the angular Einstein radius $\thetae$ and microlens parallax $\pie$, for example, OGLE-2017-BLG-0896 
\citep{Shvartzvald2019}, but the fraction of these events is small.

\citet{Han2022}, hereafter paper~I, investigated the microlensing survey data collected during 
the early phase of the KMTNet experiment with the aim of finding microlensing binaries containing 
BD companions.  The strategy applied to paper~I in finding BD events was picking out lensing 
events produced by binaries with small companion-to-primary mass ratios $q$, for example, 
$0.03\lesssim q\lesssim 0.1$.  Considering that typical Galactic lensing events are produced by 
low-mass stars \citep{Han2003}, the companions to the lenses of these events are very likely to 
be BDs.

Following the work done in Paper~I, we report  four additional BD binary-lensing events found 
from the systematic investigation of the 2018--2020 season data of the KMTNet survey.  For the 
use of a future statistical analysis on the properties of BDs based on a uniform sample, we 
consistently apply the same selection criterion as that applied in paper~I in the selection 
of BD events.

The discoveries and analyses of the BD events are presented according to the following organization. 
In Sect.~\ref{sec:two}, we mention the procedure of selecting BD binary-lens events and explain 
the observations conducted for the selected events. We describe the analysis procedure commonly 
applied to the lensing events in Sect.~~\ref{sec:three}, and detailed analyses for the individual 
events are presented in the following subsections. In Sect.~~\ref{sec:four}, we characterize the 
source stars of the events and estimate their angular Einstein radii. In Sect.~~\ref{sec:five}, we 
estimate the physical parameters of the lens systems by conducting Bayesian analyses of the events 
using the measured observables of the individual events.  A summary of the results found from the 
analyses  and conclusion are presented in Sect.~\ref{sec:six}.

\section{Event selection and observations}\label{sec:two}

The KMTNet group has conducted a microlensing survey since 2016 by observing stars lying
toward the dense Galactic bulge field with the use of three telescopes that are globally 
distributed in the Southern Hemisphere. For the searches of binary lenses possessing BD 
companions, we inspect the microlensing data acquired by the KMTNet survey during the three
seasons from 2018 to 2020. The survey in the 2020 season was partially conducted because two 
of the KMTNet telescopes were shutdown due to Covid-19 pandemic for most of that season.

We sort out binary lensing (2L1S) events with candidate BD lens companions by conducting
systematic analyses of all anomalous lensing events observed during the seasons. Anomalies
induced by planetary companions to the lenses, with companion-to-primary mass ratios of an 
order of $10^{-3}$ or less, can be, in most cases, readily identified from the characteristic 
short-term nature of the anomalies \citep{Gould1992b}. However, anomalies induced by BD 
companions, with mass ratios of an order of $10^{-2}$, usually cannot be treated as perturbations, 
and thus it is generally much more difficult to distinguish them from those induced by binary 
lenses with roughly equal-mass components. We, therefore, systematically conducted modelings 
of all anomalous lensing events detected during the seasons, and then sorted out candidate BD 
binary-lens events by applying the selection criterion of $q\lesssim 0.1$.  The total numbers 
of lensing events detected by the KMTNet survey are 2781, 3303, and 894 in the 2018, 2019, 
and 2020 seasons, respectively, and 2L1S events comprise about one tenth of the total events.  
This fraction of 2L1S events is similar to the one \citet{Shvartzvald2016} found for the 
OGLE-MOA-Wise sample (12\%).

From this procedure, we identified four 2L1S events with candidate BD companions, including 
KMT-2018-BLG-0321, KMT-2018-BLG-0885, KMT-2019-BLG-0297, and KMT-2019-BLG-0335.
We found no BD event among the events detected in the 2020 season not only because the number 
of detected lensing events during the season is relatively small but also because the data 
coverage of the individual events was sparse due to the use of a single telescope during the 
great majority of the season.  Among these events, KMT-2019-BLG-0297 was additionally 
observed by the MOA group, who labeled the event as MOA-2019-BLG-131, and we include their data 
in the analysis. For this event, we use the KMTNet ID reference following the convention of the 
microlensing community of using the ID reference of the first discovery survey

The three KMTNet telescopes are identical with a 1.6~m aperture. The sites of the individual
telescopes are the Siding Spring Observatory in Australia (KMTA), the Cerro Tololo Interamerican
Observatory in Chile (KMTC), and the South African Astronomical Observatory in South Africa
(KMTS). The telescope used for the MOA survey has an aperture of 1.8~m and is located at Mt.
John Observatory in New Zealand. The fields of view of the cameras installed on the KMTNet and
MOA telescopes are 4~deg$^2$ and 2.2~deg$^2$, respectively. Images were primarily taken in the 
$I$ band for the KMTNet survey and in the customized MOA-$R$ band for the MOA survey. For both
surveys, a minor portion of images were acquired in the $V$ band to measure the colors of the
source stars.  The OGLE survey was conducted during the 2018 and 2019 seasons, but none of the 
events reported in this work was detected by the OGLE survey.

Reductions of the images and photometry of the events were carried out using the pipelines of the
individual survey groups developed by \citet{Albrow2009} for the KMTNet group and \citet{Bond2001} 
for the MOA group.  Following the routine of \citet{Yee2012}, we readjust the error bars of each 
data set estimated by the pipelines in order that the error bars are consistent with the scatter 
of the data and $\chi^2$ per degree of freedom (dof) for each data set becomes unity.  In the 
process of readjusting error bars, we use the best-fit model by after rejecting outliers lying 
beyond a 3$\sigma$ level from the best-fit model.  The error-bar normalization is an ever-repeating 
process because once the error bars are rescaled based on a model obtained at a certain stage, the 
$\chi^2$/dof value can vary in the next modeling run as the model slightly varies from the initial 
model, and thus the value of $\chi^2$/dof can be slightly different from unity.  We note that the 
variation of the lensing parameters caused by the slight change of the  $\chi^2$/dof value is very
minor.

\section{Light curve analyses}\label{sec:three}

Under the approximation of a rectilinear relative motion between the lens and source, the light 
curve of a 2L1S event is characterized by 7 basic lensing parameters. The first three parameters 
$(t_0, u_0, \te)$ define the lens-source approach, and the individual parameters denote the time 
of the closest approach, the separation between the lens and source at that time (impact parameter), 
and the Einstein time scale. The Einstein time scale is defined as the time required for a source 
to cross the Einstein radius, that is, $\te=\thetae/\mu$,  where $\mu$ denotes the relative 
lens-source proper motion.  Another three parameters $(s, q, \alpha)$ define the binary-lens system, 
and $s$ denotes the projected separation between the lens components with masses $M_1$ and $M_2$, 
$q=M_1/M_2$ is the mass ratio, and $\alpha$ represents the angle between the direction of $\muvec$ 
and $M_1$--$M_2$, axis (source trajectory angle).  Here $\muvec$ represents the vector of the 
relative lens-source proper motion.  The parameters $u_0$ and $s$ are scaled to $\thetae$. The 
last parameter $\rho$ is defined as the ratio of the angular source radius $\theta_*$ to $\thetae$, 
that is, $\rho=\theta_*/\thetae$ (normalized source radius), and it characterizes the deformation 
of a lensing light curve by finite-source effects arising when a source crosses or approaches lens 
caustics.

Caustics represent source positions at which the lensing magnifications of a point source become 
infinity. Caustics in binary lensing vary depending on the binary parameters $s$ and $q$, and their 
topologies are classified into three categories of ``close'', ``intermediate'', and ``wide'' 
\citep{Schneider1986, Cassan2008}. A close binary induces three sets of caustics, in which one lies 
near the heavier lens component and the other two sets lie on the opposite side of the lighter lens 
component. On the other hand, a wide binary induces two sets of caustics, which lie close to the 
individual lens components. In the intermediate regime, the caustics merge together to form a single 
set of a large caustic.

Besides the basic parameters, detailed modeling of lensing light curves for a fraction of events 
requires consideration of higher-order effects caused by the deviation of the relative lens-source 
motion from rectilinear. Such a deviation is induced by two major causes, in which the first is the 
accelerated motion of an observer caused on the orbital motion of Earth, microlens-parallax effects 
\citep{Gould1992a}, and the second is the orbital motion of the binary lens, lens-orbital effects,
for example, \citet{Batista2011} and \citet{Skowron2011}.  For the consideration of these higher-order 
effects, extra parameters are required to be added in modeling. The parameters for the consideration 
of the microlens-parallax effects are $(\pien, \piee)$, which represent the north and east components 
of the microlens-parallax vector $\pivec_{\rm E}=(\pi_{\rm rel}/\thetae)(\muvec/\mu)$, respectively.  
Here $\pi_{\rm rel}={\rm AU}(D_{\rm L}^{-1}-D_{\rm S}^{-1})$ represents the relative parallax of the 
lens and source. Under the approximation of a minor change of the lens configuration by the orbital 
motion, the lens-orbital effects are described by two parameters of $(ds/dt, d\alpha/dt)$, which 
represent the change rates of the binary separation and source trajectory angle, respectively.

% Figure 1 ------------------------------------------------------
\begin{figure}[t]
\includegraphics[width=\columnwidth]{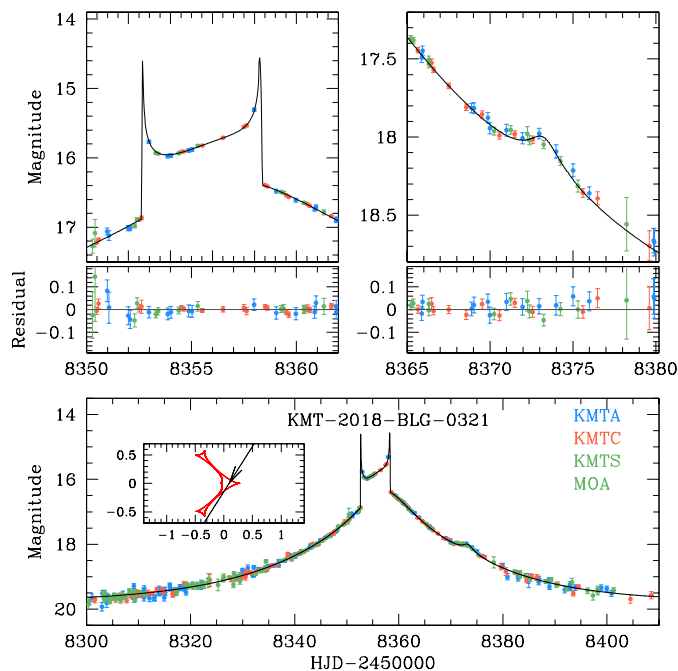}
\caption{
Light curve of KMT-2018-BLG-0321. The lower panel shows the whole view of the light curve and 
the two upper panels show the enlarged views of the two anomaly regions. The curve drawn over 
data points is the best-fit 2L1S model.  The inset in the lower panel is the lens system 
configuration, which shows the source trajectory (line with an arrow) with respect to the lens 
caustic (red figure).  Lengths are normalized to the angular Einstein radius corresponding to 
the total mass of the lens system.
}
\label{fig:one}
\end{figure}
% --------------------------------------------------------------

The analyses of the events were carried out by finding lensing solutions, representing a set 
of parameters describing the observed light curves. The searches for the lensing parameters 
were done in two steps.  In the first step, we conducted grid searches for the binary parameters 
$s$ and $q$, and for each pair of the grid parameters $s$ and $q$, we found the other parameters 
using a downhill approach based on the Markov Chain Monte Carlo (MCMC) logic.  In this stage, we 
constructed a $\chi^2$ map on the plane of the grid parameters and identified local solutions on 
the map. In the second step, we refined the individual local solutions by allowing all parameters 
to vary. If a single solution can be distinguished from the other local minima with a significant 
$\chi^2$ difference, we provide a single global solution. If the degeneracy between local solutions 
is severe, by contrast, we present all local solutions with the explanations on the causes of the 
degeneracy. In the following subsections, we present the analyses of the individual events.

\subsection{KMT-2018-BLG-0321}\label{sec:three-one}

The source of the lensing event KMT-2018-BLG-0321 lies in the Galactic bulge field with Equatorial 
coordinates (RA, DEC)$_{\rm J2000}= ($17:41:36.41, -22:09:52.88), which correspond to the Galactic 
coordinates $(l, b) = (5^\circ\hskip-2pt .301, 4^\circ\hskip-2pt .315)$.  The event was detected by 
the KMTNet survey on 2018 July 21 (${\rm HJD}^\prime \equiv {\rm HJD}-2450000 = 8320.46$), when the 
source was brighter by $\Delta I\sim 1.1$ magnitude than the baseline of $I_{\rm base}=18.23$ using 
the AlertFinder algorithm of the KMTNet survey \citep{Kim2018b}. The source of the event lies in the 
KMTNet BLG20 field, toward which observations were conducted with a 2.5~hr cadence.

The lensing light curve of KMT-2018-BLG-0321 is presented in Figure~\ref{fig:one}. It shows 
that the light curve exhibits deviations from the smooth and symmetric form of a single-lens 
single-source (1L1S) event. The deviations are characterized by three major anomaly features, 
including the two spike features appearing around the peak of the light curve at ${\rm HJD}^\prime 
\sim 8352.6$ and 8358.2, and the weak bump appearing on the falling side of the light curve at 
${\rm HJD}^\prime\sim 8373.1$. From their shapes, the two spike features are likely to be produced 
by the source crossings over folds of a binary caustic, and the bump feature is likely to be 
generated by the source approach to a cusp of the caustic

% Table 1 ------------------------------------------------
\begin{table}[t]
\small
%\centering
\caption{Model parameters of KMT-2018-BLG-0321\label{table:one}}
\begin{tabular*}{\columnwidth}{@{\extracolsep{\fill}}lcccc}
\hline\hline
\multicolumn{1}{c}{Parameter}    &
\multicolumn{1}{c}{Value }       \\
\hline
$\chi^2$/dof            &   $721.4/716            $      \\
$t_0$ (HJD$^\prime$)    &   $8355.527 \pm 0.019   $      \\
$u_0$                   &   $0.076  \pm 0.001     $      \\
$\te$ (days)            &   $25.57 \pm 0.16       $      \\
$s$                     &   $0.799 \pm 0.002      $      \\
$q$                     &   $0.103 \pm 0.002      $      \\
$\alpha$ (rad)          &   $5.276 \pm 0.005      $      \\
$\rho$ ($10^{-3}$)      &   $< 2.5                $      \\
\hline
\end{tabular*}
\tablefoot{ ${\rm HJD}^\prime = {\rm HJD}- 2450000$.  }
\end{table}
% --------------------------------------------------------

From the detailed modeling of the observed light curve, we found that the event was generated by 
a binary lens with a small mass ratio between the lens components. We found a unique solution 
without any degeneracy, and the estimated binary parameters are $(s, q)\sim (0.8, 0.1)$. The exact 
values of the lensing parameters are listed in Table~\ref{table:one} together with the value of 
$\chi^2/{\rm dof}$. Neither of the caustic-crossing features was resolved, and only an upper limit 
on $\rho$ could be constrained. It was found that secure measurements of the higher-order lensing 
parameters were difficult due to the moderate time scale, $\te \sim 26$~days, of the event.

The inset in the bottom panel of Figure~\ref{fig:one} shows the lens system configuration of the 
event. The caustic is at the boundary between the close and intermediate regimes, and the two 
peripheral caustics are connected with the central caustic by slim bridges.  The best-fit model 
indicates that the two caustic spikes were produced when the source entered and exited the central 
caustic, and the weak bump on the falling side of the light curve was produced when the source 
approached close to one of the two peripheral caustics.

\subsection{KMT-2018-BLG-0885}\label{sec:three-two}

The lensing event KMT-2018-BLG-0885 occurred on a source lying at (RA, DEC)$_{\rm J2000}= ($17:55:51.92, 
-28:30:48.10), $(l, b) = (1^\circ\hskip-2pt .513, -1^\circ\hskip-2pt .713)$.  The event occurred before 
the full operation of the KMTNet AlertFinder system, and it was identified from the inspection of the 
data after the bulge season was over by the KMTNet EventFinder system \citep{Kim2018a}.  The source of 
the event lies in the two overlapping KMTNet fields of BLG02 and BLG42, each of which was monitored with 
a 0.5~hr cadence, and thus with a combined cadence of 0.25~hr.

% Figure 2 ------------------------------------------------------
\begin{figure}[t]
\includegraphics[width=\columnwidth]{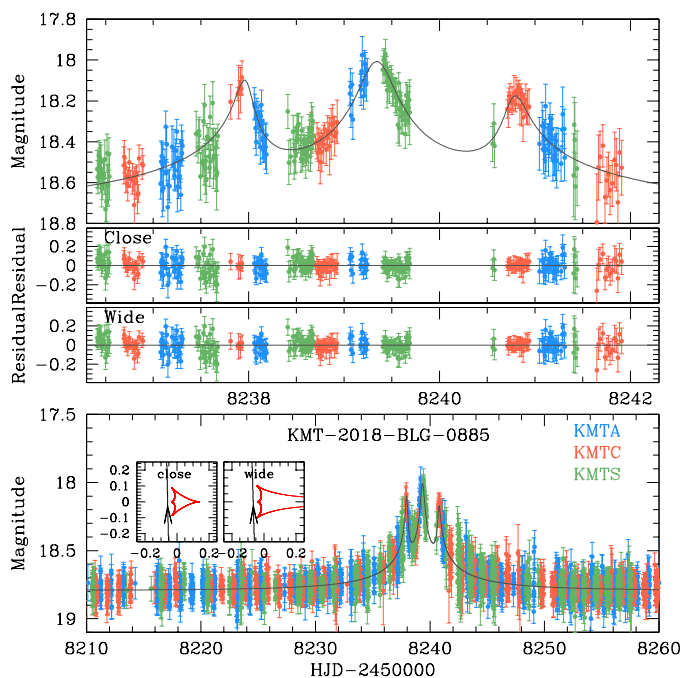}
\caption{
Light curve of KMT-2018-BLG-0885.  Notations are the same as those in Fig.~\ref{fig:one}. For 
this event, there exist two degeneration solutions (``close'' and ``wide'' solutions), and two 
sets of lens system configuration, presented in the bottom panel, are presented.
}
\label{fig:two}
\end{figure}
% --------------------------------------------------------------

The light curve of the event is shown in Figure~\ref{fig:two}. It is characterized by three consecutive 
bumps appearing around the peak with roughly 2-day gaps between each consecutive pair of bumps.  The 
deviation pattern of these bumps are smooth, suggesting that they were produced by successive approaches 
of the source close to three cusps of a caustic. For this event, the low mass ratio between the lens 
components was expected to some extent, because triple-bump anomalies can be produced when a caustic 
is skewed and its cusps lie on one side of the caustic, for example, the second microlensing planet, 
OGLE-2005-BLG-071Lb \citep{Udalski2005}.

From detailed modeling of the light curve, we identified two solutions, in which one is in the 
close binary regime and the other is in the wide binary regime. The binary parameters are 
$(s, q)_{\rm close}\sim (0.62, 0.096)$ for the close solution and $(s, q)_{\rm wide}\sim 
(1.53, 0.102)$ for the wide solution. The fact that the binary separations of the close and wide 
solutions approximately follow the relation $s_{\rm close}\sim 1/s_{\rm wide}$ indicates that the 
degeneracy between the solutions is caused by the close--wide degeneracy \citep{Griest1998, 
Dominik1999, An2005}. We present the full lensing parameters of both solutions in 
Table~\ref{table:two}. As expected from the anomaly pattern, the companion-to-primary mass ratio, 
$q\sim 0.1$, of the lens is small, suggesting that the companion to the lens is likely to be a BD. 
It was found that the close solution yields a slightly better fit to the data over the wide solution, 
but the difference between the fits of the two solutions is small with $\Delta\chi^2=6.1$.  In 
Figure~\ref{fig:two}, we draw the model curve of the close solution, and present the residuals from 
the close and wide solutions in the region of the anomalies.

% Table 2 ------------------------------------------------
\begin{table}[t]
\small
%\centering
\caption{Model parameters of KMT-2018-BLG-0885\label{table:two}}
\begin{tabular*}{\columnwidth}{@{\extracolsep{\fill}}lcccc}
\hline\hline
\multicolumn{1}{c}{Parameter}   &
\multicolumn{1}{c}{Close}       &
\multicolumn{1}{c}{Wide}        \\
\hline
$\chi^2$/dof            &   $9735.8/9706         $    &   $9741.9/9706         $      \\
$t_0$ (HJD$^\prime$)    &   $8239.331 \pm 0.012  $    &   $8239.329 \pm 0.014  $      \\
$u_0$                   &   $0.058 \pm  0.004    $    &   $0.067 \pm 0.005     $      \\
$\te$ (days)            &   $12.14 \pm  0.72     $    &   $11.72 \pm 0.75      $      \\
$s$                     &   $0.618 \pm  0.007    $    &   $1.528 \pm 0.027     $      \\
$q$                     &   $0.096 \pm  0.006    $    &   $0.102 \pm 0.007     $      \\
$\alpha$ (rad)          &   $1.549 \pm  0.006    $    &   $1.543 \pm 0.012     $      \\
$\rho$ ($10^{-3}$)      &   $< 10                $    &   $< 10                $      \\
\hline
\end{tabular*}
%\tablefoot{ ${\rm HJD}^\prime = {\rm HJD}- 2450000$.  }
\end{table}
% --------------------------------------------------------

The two insets in the bottom panel of Figure~\ref{fig:two} show the configurations of the close 
(left inset) and wide (right inset) solutions. Each configuration shows that the three bumps were 
produced by the source passage close to the three protruding cusps of a caustic generated by a binary 
lens with a low mass ratio. The normalized source radius could not be tightly constrained and only 
its upper limit, $\rho_{\rm max}\sim 0.01$, can be set.

\subsection{KMT-2019-BLG-0297}\label{sec:three-three}

The source coordinates of the event KMT-2019-BLG-0297 are (RA, DEC)$_{\rm J2000}= ($18:00:15.19, 
-28:57:55.01), $(l, b) = (1^\circ\hskip-2pt .602, -2^\circ\hskip-2pt .772)$.  The source lies in 
the two overlapping KMTNet prime fields of BLG03 and BLG43, toward which the event was observed 
with a 0.25~hr combined cadence. In the 2019 season, the AlertFinder system was operational, and 
the event was detected in its early stage on April 5 (${\rm HJD}^\prime \sim 8578.8$), when the 
event was magnified by $\Delta I\sim 1.1$ magnitude from the baseline of $I_{\rm base}=19.90$. The 
event was independently detected by the MOA survey three days after the KMTNet alert.

% Figure 3 ------------------------------------------------------
\begin{figure}[t]
\includegraphics[width=\columnwidth]{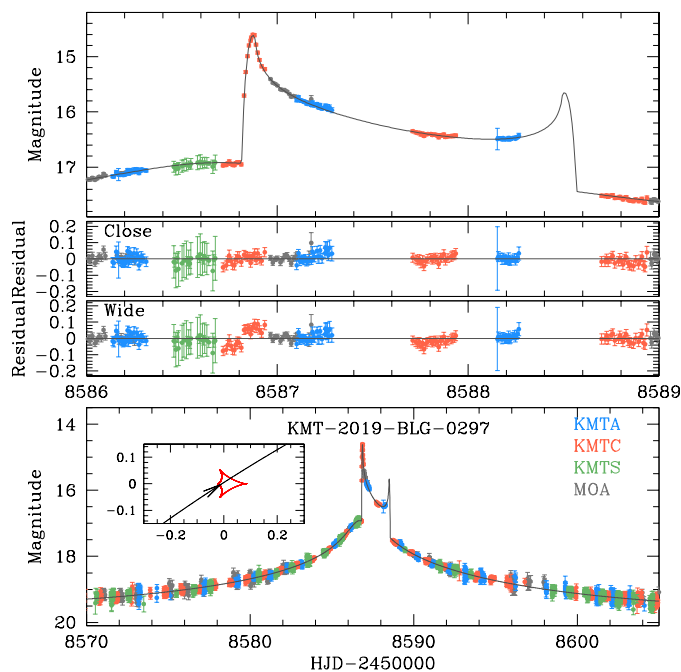}
\caption{
Light curve of KMT-2019-BLG-0297.  Notations are same as those in Fig.~\ref{fig:one}. The model 
curve drawn over the data points is that of the close solution found considering the higher-order 
effects
}
\label{fig:three}
\end{figure}
% --------------------------------------------------------------

% Table 3 ------------------------------------------------
\begin{table*}[t]
\small
%\centering
\caption{Model parameters of KMT-2019-BLG-0297\label{table:three}}
\begin{tabular}{lccccc}
%\begin{tabular}{\columnwidth}{@{\extracolsep{\fill}}lllcc}
\hline\hline
\multicolumn{1}{c}{Parameter}         &
\multicolumn{2}{c}{Standard}          &
\multicolumn{1}{c}{Higher order}      \\
\multicolumn{1}{c}{}         &
\multicolumn{1}{c}{Close}          &
\multicolumn{1}{c}{Wide}          &
\multicolumn{1}{c}{(close)}      \\
\hline
$\chi^2$/dof               &   $4876.2/4876        $     &  $5062.2/4876        $   &  $4791.9/4872        $      \\
$t_0$ (HJD$^\prime$)       &   $8587.370 \pm 0.006 $     &  $8587.593 \pm  0.006$   &  $8587.413 \pm 0.008 $      \\
$u_0$                      &   $0.0031 \pm 0.0003  $     &  $0.0071 \pm 0 .0002 $   &  $0.0042 \pm 0.0003  $      \\
$\te$ (days)               &   $31.62 \pm 0.31     $     &  $41.36 \pm 0. 79    $   &  $35.33 \pm 0.77     $      \\
$s$                        &   $0.563 \pm 0.002    $     &  $2.718 \pm 0. 021   $   &  $0.528 \pm 0.004    $      \\
$q$                        &   $0.081 \pm 0.002    $     &  $0.188 \pm 0. 007   $   &  $0.086 \pm 0.003    $      \\
$\alpha$ (rad)             &   $2.553 \pm 0.005    $     &  $2.606 \pm 0. 004   $   &  $2.568 \pm 0.005    $      \\
$\rho$ ($10^{-3}$)         &   $1.195 \pm 0.024    $     &  $0.837 \pm 0. 021   $   &  $1.046 \pm 0.029    $      \\
$\pien$                    &   --                        &  --                      &  $0.50 \pm 0.72      $      \\
$\piee$                    &   --                        &  --                      &  $-0.29 \pm 0.07     $      \\
$ds/dt$ (yr$^{-1}$)        &   --                        &  --                      &  $-0.87 \pm 0.14     $      \\
$d\alpha/dt$ (yr$^{-1}$)   &   --                        &  --                      &  $-0.61 \pm 1.4      $      \\
\hline
\end{tabular}
%\tablefoot{ $(V-I)_{0,{\rm RGC}}=1.06$  }
\end{table*}
% --------------------------------------------------------

The lensing light curve of the event constructed with the combination of the KMTNet and MOA 
data is shown in Figure~\ref{fig:three}. It is characterized by a central anomaly that lasted for 
about two days. The main feature of the anomaly is the sharp spike centered at ${\rm HJD}^\prime 
\sim 8586.9$ caused by a caustic crossing. Binary caustics form closed curves, and thus caustic 
crossings occur in pairs, that is, when the source enters and exits the caustic. Then, there should 
be an additional caustic spike, although the data did not cover it. From the curvature of the 
U-shape pattern after the first caustic spike, it is expected that the second caustic spike occurred 
at around ${\rm HJD}^\prime \sim 8588.5$.

% Figure 4 ------------------------------------------------------
\begin{figure}[t]
\includegraphics[width=\columnwidth]{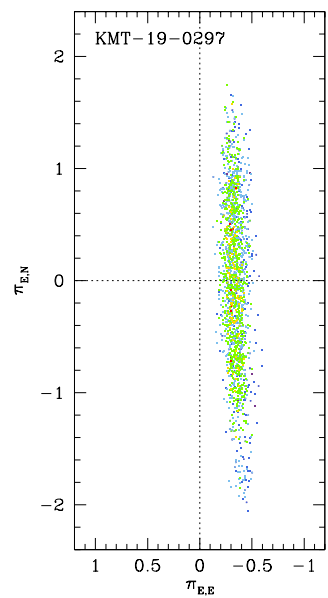}
\caption{
Scatter plot of MCMC points on the $(\piee, \pien)$ plane obtained from modeling the light 
curve of KMT-2019-BLG-0297 with the consideration of higher-order effects.  The color 
coding represent points with $\leq 1\sigma$ (red), $\leq 2\sigma$ (yellow), $\leq 3\sigma$ 
(green), $\leq 4\sigma$ (cyan), and $\leq 5\sigma$ (blue).
}
\label{fig:four}
\end{figure}
% --------------------------------------------------------------

From the detailed modeling of the light curve, it was found that the anomaly in the light 
curve of the event KMT-2019-BLG-0297 was generated by a close binary ($s\sim 0.56$) with a 
low mass ratio $(q\sim 0.08)$ between the lens components. The full lensing parameters of 
the solution are presented in Table~\ref{table:three}. We find a weak local minimum of a wide 
binary lens with $(s, q)\sim (2.72, 0.19)$, but its fit is worse than that of the close 
solution by $\Delta\chi^2 =186.0$.

With relatively high-precision data in the wings of the light curve, we test whether the fit 
further improves with the consideration of higher-order effects. For this check, we conduct an 
additional modeling considering both the microlens-parallax and lens-orbital effects. It was 
found that the model considering higher-order effects substantially improved the fit by 
$\Delta\chi^2 =84.3$. The lensing parameters of the higher-order solution are listed in 
Table~\ref{table:three}, and the model curve and residual around the anomaly region are shown 
in Figure~\ref{fig:three}.  Although turned down, we present the residual of the wide solution 
for the comparison with the close solution.  We note that the variations of the basic lensing 
parameters with the inclusion of higher-order effects are minor. In Figure~\ref{fig:four}, we 
present the scatter plot of points in the MCMC chain on the $(\pien, \piee)$ plane.  The plot 
shows that the east component of the microlens-parallax vector is relatively well constrained, 
although the  north component is not securely measured.  As will be discussed in Sect.~\ref{sec:five}, 
the measurement of the microlens parallax is important because $\pie$ provides an extra constraint 
on the physical parameters of the lens. For KMT-2019-BLG-0297, the angular Einstein radius, which 
is another observable related to the physical lens parameters, can also be constrained because the 
caustic entrance was densely resolved by the KMTC data, and this yields the normalized source radius 
$\rho$, from which the angular Einstein radius is determined as
\begin{equation}
\thetae={\theta_*\over \rho}. 
\label{eq1}
\end{equation}
More details about the $\thetae$ determination are discussed in Sect.~\ref{sec:four}.

% Figure 5 ------------------------------------------------------
\begin{figure}[t]
\includegraphics[width=\columnwidth]{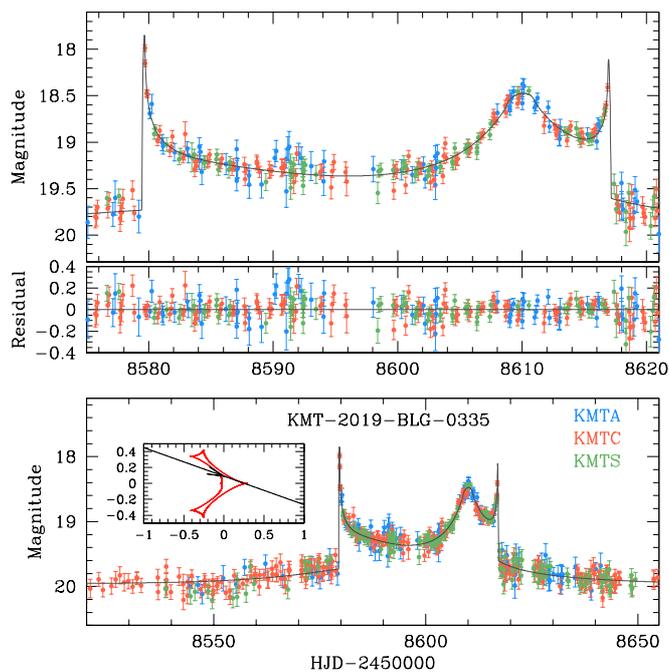}
\caption{
Light curve of KMT-2019-BLG-0335.  Notations are same as those in Fig.~\ref{fig:one}.
}
\label{fig:five}
\end{figure}
% --------------------------------------------------------------

% Figure 6 ------------------------------------------------------
\begin{figure*}[t]
\centering
\includegraphics[width=11.8cm]{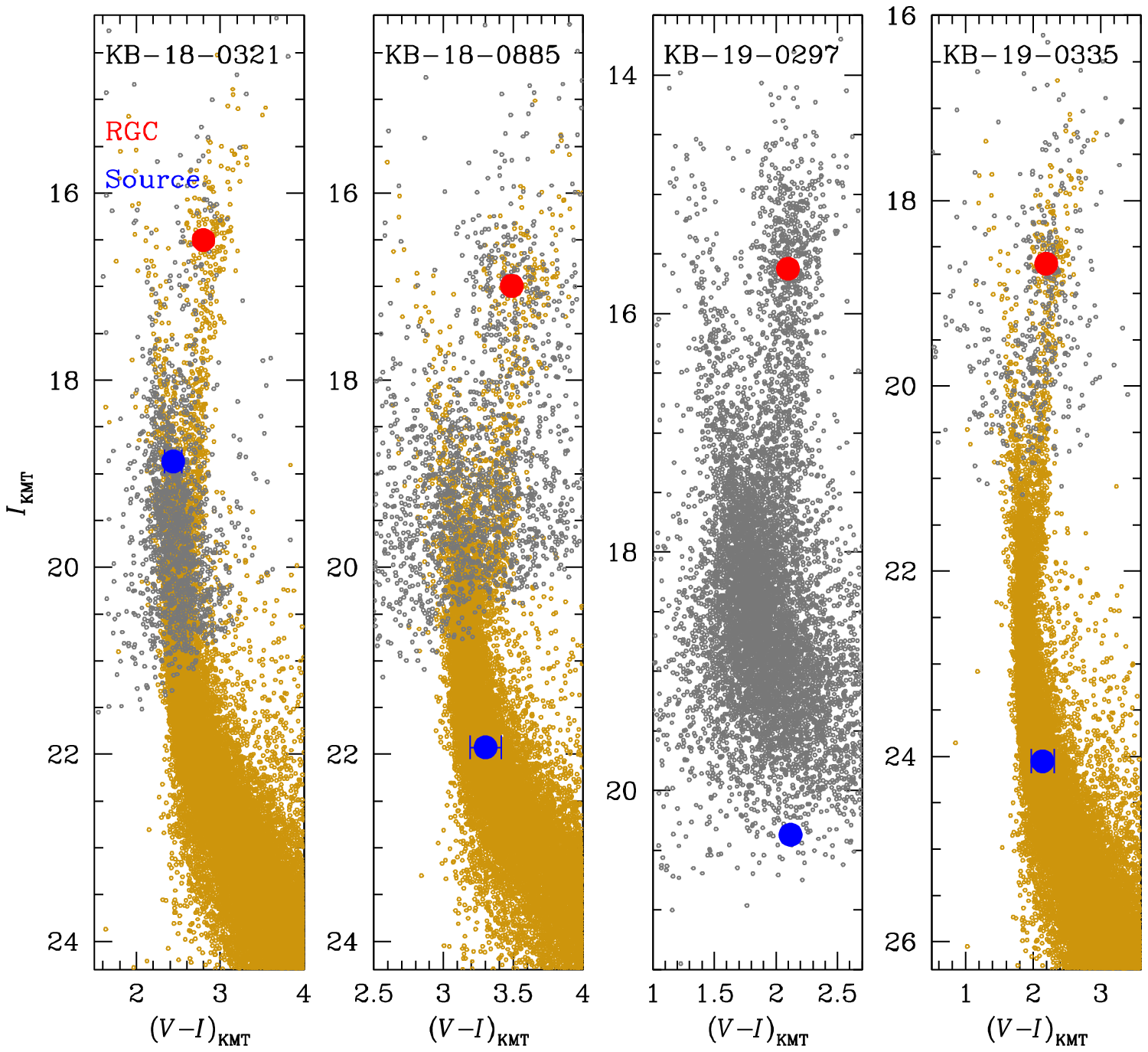}
\caption{
Source locations (blue filled dot) with respect to the centroids of red giant clump (RGC, 
red filled dot) in the color-magnitude diagrams. For the three events KMT-2018-BLG-0321, 
KMT-2018-BLG-0885, and KMT-2019-BLG-0335, the CMDs are constructed by combining those from 
KMTC (grey dots) and HST (brown dots) observations.
}
\label{fig:six}
\end{figure*}
% --------------------------------------------------------------

\subsection{KMT-2019-BLG-0335}\label{sec:three-four}

The source of the lensing event KMT-2019-BLG-0335 lies at (RA, DEC)$_{\rm J2000}= ($17:31:21.63, 
-29:35:26.48), $(l, b) = (-2^\circ\hskip-2pt .215, 2^\circ\hskip-2pt .279)$.  The source position 
corresponds to the KMTNet BLG11 field, toward which observations were conducted with a 2.5~hr 
cadence. The lensing magnification of the source flux began before the start of the 2019 observation 
season, and the event was detected on 2019 April 9 (${\rm HJD}^\prime =8582.6$) after the event went 
through a substantial magnification by the caustic crossing of the source.

% Table 4 ------------------------------------------------
\begin{table}[t]
\small
%\centering
\caption{Model parameters of KMT-2019-BLG-0335\label{table:four}}
\begin{tabular*}{\columnwidth}{@{\extracolsep{\fill}}lcccc}
\hline\hline
\multicolumn{1}{c}{Parameter}    &
\multicolumn{1}{c}{Value }       \\
\hline
$\chi^2$/dof            &   $819.9/823           $     \\
$t_0$ (HJD$^\prime$)    &   $8589.772 \pm 0.346  $     \\
$u_0$                   &   $0.087 \pm 0.004     $     \\
$\te$ (days)            &   $109.22 \pm 3.48     $     \\
$s$                     &   $0.856 \pm 0.006     $     \\
$q$                     &   $0.064 \pm 0.004     $     \\
$\alpha$ (rad)          &   $3.486 \pm 0.008     $     \\
$\rho$ ($10^{-3}$)      &   $0.57 \pm 0.15       $     \\
\hline
\end{tabular*}
%\tablefoot{ ${\rm HJD}^\prime = {\rm HJD}- 2450000$.  }
\end{table}
% --------------------------------------------------------

Figure~\ref{fig:five} shows the lensing light curve of KMT-2019-BLG-0335. It is characterized by
three distinctive anomaly features: the two spikes at ${\rm HJD}^\prime \sim 8579.0$ and 8616.3 
and the bump centered at ${\rm HJD}^\prime \sim 8610$ appearing between the caustic spikes.  
From the characteristic pattern, the spike features are likely to result from caustic 
crossings. From the location of the bump appearing in the U-shape region between the 
caustic-crossing features, it is expected that the bump was produced by the asymptotic approach 
of the source close to a fold of a caustic as the source proceeded inside the caustic.  
Although the rising part of the first spike feature and the falling part of the second spike 
feature were not covered by the data, the pattern of the caustic-crossing features can be well 
delineated by the data just after the first spike and before the second spike.

Detailed modeling of the light curve yields a unique solution with binary parameters of 
$(s, q)\sim (0.86, 0.064)$, indicating that the anomaly features were produced by a close 
binary with a low-mass companion. We list the full lensing parameters in Table~\ref{table:four}, 
and the model curve and residual around the anomaly region are presented in Figure~\ref{fig:five}. 
The time scale of the event, $\te \sim 109$~days, is fairly long, but it is difficult to constrain 
the higher-order lensing parameters due to the substantial photometric errors of the data caused 
by the faintness of the source. The normalized source radius is measured, although its uncertainty 
is fairly big due to the incomplete coverage of the caustic crossings.

The inset in the bottom panel of Figure~\ref{fig:five} shows the configuration of the lens system. 
The caustic is similar to that of KMT-2018-BLG-0321 with the two peripheral caustics connected with 
the central caustic by narrow bridges, indicating that the binary is at the boundary between the 
close and intermediate regimes. The source entered the upper left side of the central caustic, 
passed along the upper right fold of the caustic, and exited the lower right side of the caustic. 
The caustic entrance and exit produced the spike features, and the bump was generated by the source 
approach to the caustic fold.

% Table 5 ------------------------------------------------
\begin{table*}[t]
\small
%\centering
\caption{Source property\label{table:five}}
\begin{tabular}{llcccc}
%\begin{tabular}{\columnwidth}{@{\extracolsep{\fill}}lllcc}
\hline\hline
\multicolumn{1}{c}{Events}    &
\multicolumn{1}{c}{$(V-I, I)_{\rm S}$}    &
\multicolumn{1}{c}{$(V-I, I)_{\rm RGC}$}    &
\multicolumn{1}{c}{$I_{0,{\rm RGC}}$}      &
\multicolumn{1}{c}{$(V-I, I)_{0,{\rm S}}$}    &
\multicolumn{1}{c}{$\theta_*$ ($\mu$as)}       \\
\hline
KMT-2018-BLG-0321  & $(2.441\pm 0.103, 18.868\pm 0.029)$  &  $(2.800, 16.500)$    &  16.500 &  $(0.701\pm 0.103, 16.663\pm 0.029)$   &  $1.44\pm 0.18$ \\
KMT-2018-BLG-0885  & $(3.301\pm 0.112, 21.925\pm 0.069)$  &  $(3.490, 16.990)$    &  14.384 &  $(0.871\pm 0.112, 19.319\pm 0.069)$   &  $0.52\pm 0.07$ \\
KMT-2019-BLG-0297  & $(2.119\pm 0.007, 20.371\pm 0.002)$  &  $(2.099, 15.624)$    &  14.382 &  $(1.079\pm 0.007, 19.129\pm 0.002)$   &  $0.71\pm 0.05$ \\
KMT-2019-BLG-0335  & $(2.139\pm 0.171, 24.051\pm 0.028)$  &  $(2.192, 18.680)$    &  14.396 &  $(1.007\pm 0.171, 19.766\pm 0.028)$   &  $0.49\pm 0.09$ \\
%KMT-2019-BLG-3301  & $(1.954\pm 0.010, 19.455\pm 0.006)$  &  $(2.219, 15.597)$    &  14.387 &  $(0.795\pm 0.180, 18.245\pm 0.008)$   &  $0.78\pm 0.05$ \\
\hline
\end{tabular}
\tablefoot{ $(V-I)_{0,{\rm RGC}}=1.06$  }
\end{table*}

\section{Source stars and Einstein radii}\label{sec:four}

In this section, we specify the source stars of the individual lensing events and estimate angular 
Einstein radii for the events with measured normalized source radii. For each event, the source is 
specified by measuring its reddening and extinction-corrected (de-reddened) color and magnitude. 
The measured source color and magnitude are used to deduce the angular source radius, from which 
the angular Einstein radius is estimated from the relation in Equation~(\ref{eq1}).

Figure~\ref{fig:six} shows the source locations of the individual lensing events in the instrumental
color-magnitude diagrams (CMDs) of stars lying adjacent to the source stars constructed from pyDIA 
\citep{Albrow2017} photometry of the KMTC images. The $I$- and $V$-band magnitudes of each source 
were estimated from the regression of the light curve data measured using the same pyDIA photometry 
with respect to the lensing magnification. For the three events KMT-2018-BLG-0321, KMT-2018-BLG-0885, 
and KMT-2019-BLG-0335, the $V$-band source magnitudes could not be securely measured due to the poor 
quality of the $V$-band data, although their $I$-band magnitudes were measured. In these cases, we 
first combine the two sets of CMDs, one constructed from the pyDIA photometry of stars in the KMTC 
image and the other for stars in the Baade's window observed with the use of the Hubble Space Telescope 
\citep{Holtzman1998}, align the two CMDs using the centroids of the red giant clump (RGC) in the 
individual CMDs, and then estimate the source color as the median values for stars in the main-sequence 
branch of the HST CMD with $I$-band magnitude offsets from the RGC centroid corresponding to the measured 
values. The estimated instrumental colors and magnitudes of the source stars, $(V-I, I)_{\rm S}$, and RGC 
centroids, $(V-I, I)_{\rm RGC}$, for the individual events are listed in Table~\ref{table:five}.

% Figure 7 ------------------------------------------------------
\begin{figure}[t]
\includegraphics[width=\columnwidth]{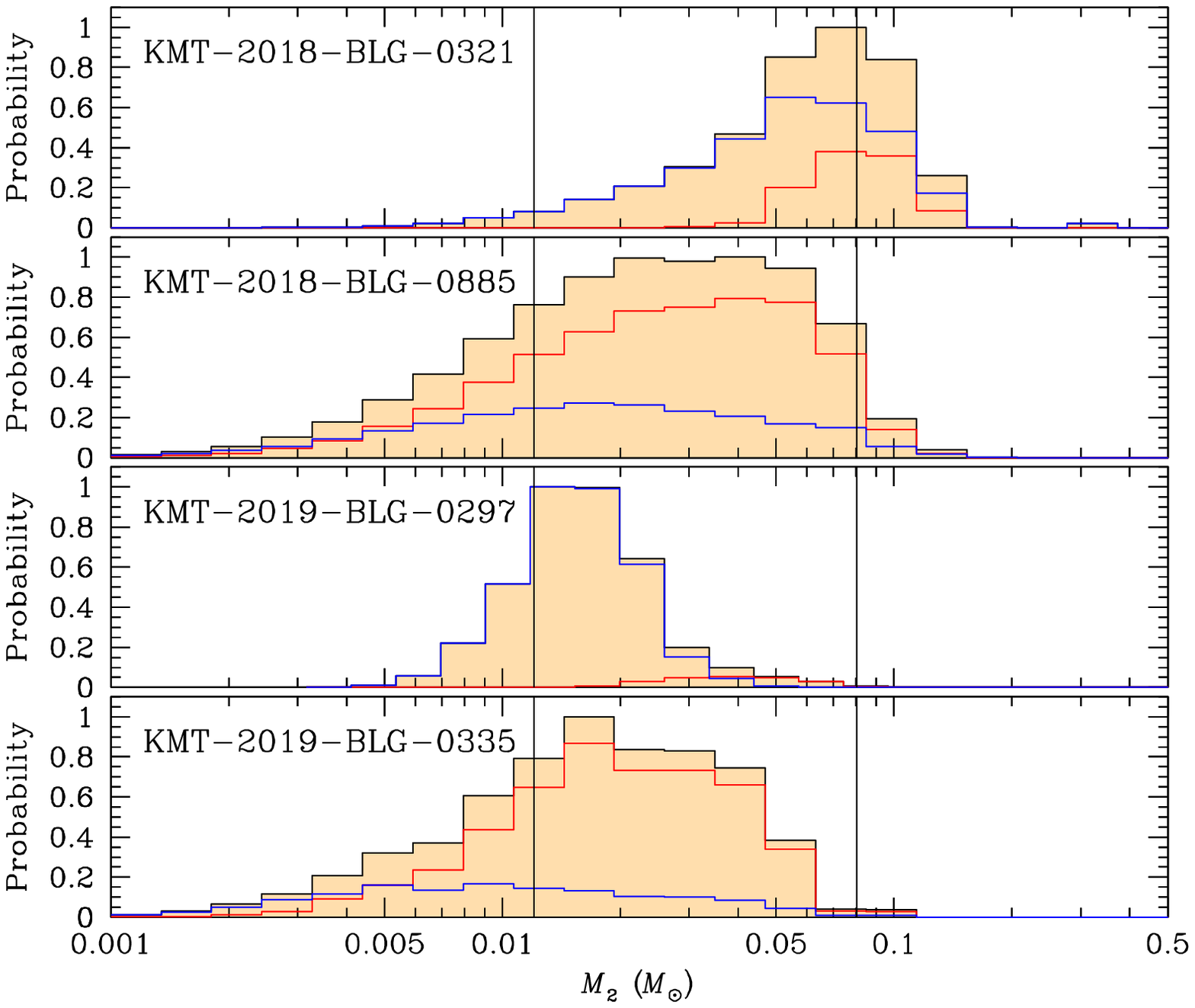}
\caption{
Bayesian posteriors for the masses of the lens companions.  The blue and red curves in each panel 
indicate the contributions by the disk and bulge lens populations, respectively, and the black curve 
is the sum of the contributions. The two vertical lines at $M_2=0.012~M_\odot$ ($\sim 12~M_{\rm J}$) 
and $0.08~M_\odot$ indicate the boundaries between planetary, BD, and stellar lens populations.
}
\label{fig:seven}
\end{figure}
% --------------------------------------------------------------

% Table 6 ------------------------------------------------
\begin{table}[t]
\small
%\centering
\caption{Einstein radius and proper motion\label{table:six}}
\begin{tabular*}{\columnwidth}{@{\extracolsep{\fill}}lllcc}
\hline\hline
\multicolumn{1}{c}{Event}                        &
\multicolumn{1}{c}{$\thetae$ (mas)}              &
\multicolumn{1}{c}{$\mu$ (mas~yr$^{-1}$) }       \\
\hline
KMT-2018-BLG-0321    &  $> 0.57         $    &  $> 2.4            $  \\
KMT-2018-BLG-0885    &  $> 0.05         $    &  $> 1.6            $  \\
KMT-2019-BLG-0297    &  $0.60 \pm 0.04  $    &  $6.93 \pm 0.51    $  \\
KMT-2019-BLG-0335    &  $0.19 \pm 0.07  $    &  $2.87 \pm 1.16    $  \\
%KMT-2019-BLG-3301    &  $0.72 \pm 0.09  $    &  $5.02 \pm 0.63    $  \\
\hline
\end{tabular*}
%\tablefoot{ ${\rm HJD}^\prime = {\rm HJD}- 2450000$.  }
\end{table}
% --------------------------------------------------------

For the calibration of the source colors and magnitudes, we use the RGC centroid, for which its 
de-reddened values $(V-I, I)_{0,{\rm RGC}}$ are well defined \citep{Bensby2013, Nataf2013}, as a 
reference \citep{Yoo2004}.  By measuring the offsets in color and magnitude, $\Delta (V-I, I)$, 
of the source star from those of the RGC centroid, the de-reddened values are estimated as 
$(V-I, I)_{0, {\rm S}} = (V-I, I)_{0,{\rm RGC}} + \Delta (V-I, I)$. The estimated de-reddened 
source colors and magnitudes of the individual events are listed in Table~\ref{table:five}. 
According to the estimated de-reddened colors and magnitudes, it is found that the source of 
KMT-2018-BLG-0321 is a G-type turnoff star, and those of the other events are main-sequence stars 
with spectral types ranging from late G to early K.

The angular radii of the source stars were deduced from their measured colors and magnitudes.  For 
this, we first converted $V-I$ color into $V-K$ color using the \citet{Bessell1988} relation, and 
then estimated the angular source radius using the \citet{Kervella2004} relation between $(V-K, I)$ 
and $\theta_*$. For the events with measured normalized source radii, the angular Einstein radii were 
estimated using the relation in Equation~(\ref{eq1}). The estimated angular radii of the source stars 
and Einstein rings of the individual events are listed in Table~\ref{table:six}. Also listed are the 
relative proper motions between the lens and source estimated by $\mu =\thetae/\te$. In the cases of 
the events KMT-2018-BLG-0321 and KMT-2018-BLG-0885, for which only the upper limits of $\rho$ are 
constrained, we list the lower limits of $\thetae$ and $\mu$.

% Figure 8 ------------------------------------------------------
\begin{figure}[t]
\includegraphics[width=\columnwidth]{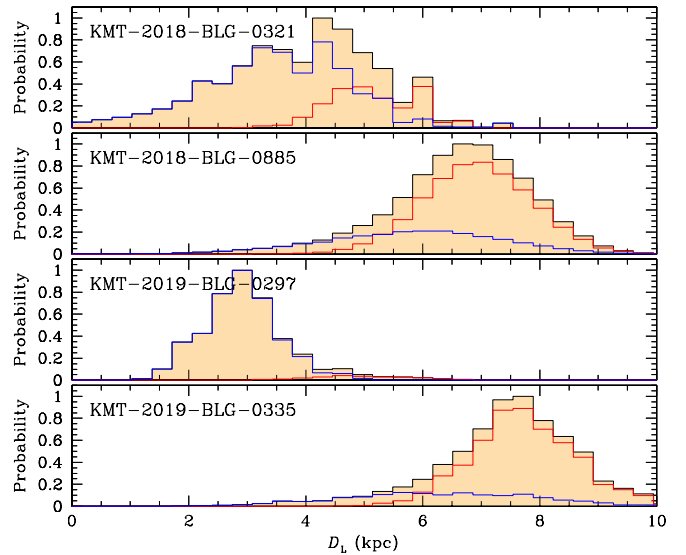}
\caption{
Bayesian posteriors for the distances of the lens systems.
Notations are the same as those in Fig.~\ref{fig:seven}.
}
\label{fig:eight}
\end{figure}
% --------------------------------------------------------------

% Table 7 ------------------------------------------------
\begin{table*}[t]
%\footnotesize
%\small
%\centering
\caption{Physical lens parameters \label{table:seven}}
\begin{tabular}{lllllrrrrr}
%\begin{tabular}{\columnwidth}{@{\extracolsep{\fill}}lllcc}
\hline\hline
\multicolumn{1}{c}{Events}              &
\multicolumn{1}{c}{$M_1$ ($M_\odot$)}   &
\multicolumn{1}{c}{$M_2$ ($M_\odot$)}   &
\multicolumn{1}{c}{$\dl$ (kpc)}         &
\multicolumn{1}{c}{$a_\perp$ (AU)}      \\
%\multicolumn{1}{c}{}                  &
%\multicolumn{1}{c}{($M_\odot$)}       &
%\multicolumn{1}{c}{($M_\odot$)}       &
%\multicolumn{1}{c}{(kpc)}             &
%\multicolumn{1}{c}{(AU)}              \\
\hline
KMT-2018-BLG-0321           &  $0.75^{+0.33}_{-0.29} $  &  $0.072^{+0.042}_{-0.038}$  &  $4.23^{+1.15}_{-1.57}$  & $2.54^{+0.69}_{-0.95} $  \\ [0.8ex]
KMT-2018-BLG-0885 (close)   &  $0.29^{+0.38}_{-0.18} $  &  $0.028^{+0.037}_{-0.018}$  &  $6.89^{+1.06}_{-1.24}$  & $1.07^{+0.17}_{-0.19} $  \\ [0.8ex]
\hskip93pt        (wide)    &   --                      &  $0.030^{+0.039}_{-0.019}$  &   --                     & $2.64^{+0.41}_{-0.47} $  \\ [0.8ex]
KMT-2019-BLG-0297           &  $0.21^{+0.09}_{-0.07} $  &  $0.018^{+0.008}_{-0.007}$  &  $3.09^{+0.76}_{-0.68}$  & $0.96^{+0.24}_{-0.21} $  \\ [0.8ex]
KMT-2019-BLG-0335           &  $0.32^{+0.35}_{-0.19} $  &  $0.020^{+0.022}_{-0.012}$  &  $7.71^{+0.99}_{-1.22}$  & $1.64^{+0.21}_{-0.26} $  \\ [0.8ex]
%KMT-2019-BLG-3301           &  $0.78^{+0.40}_{-0.36} $  &  $0.034^{+0.018}_{-0.016}$  &  $5.06^{+1.18}_{-1.43}$  & $1.41^{+0.27}_{-0.32} $  \\ [0.8ex]
\hline
\end{tabular}
%\tablefoot{ $(V-I)_{0,{\rm RGC}}=1.06$  }
\end{table*}
% --------------------------------------------------------

\section{Physical lens properties}\label{sec:five}

The basic lensing observable constraining the physical parameters of the lens mass $M$ and
distance to the lens $\dl$ is the Einstein time scale $\te$, which is related to the physical 
parameters by
\begin{equation}
\te = {\thetae\over \mu};\qquad
\thetae = (\kappa M \pi_{\rm rel})^{1/2},
\label{eq2}
\end{equation}
where $\kappa=4G/(c^2{\rm AU})$. 
Besides this observable, the lens mass and distance can be additionally constrained by measuring
the extra observables of $\pie$ and $\thetae$. If both of these extra observables are simultaneously
measured, the physical lens parameters can be uniquely determined by the relation
\begin{equation}
M={\thetae\over \kappa \pie};\qquad
\dl = { {\rm AU} \over \pie\thetae + \pi_{\rm S}},
\label{eq3}
\end{equation}
where $\pi_{\rm S}={\rm AU}/D_{\rm S}$ is the parallax of the source, and $D_{\rm S}$ denotes the 
distance to the source \citep{Gould2000}. For KMT-2019-BLG-0297, both of these extra parameters 
are measured, but the uncertainty of the measured microlens parallax is large.  For the event 
KMT-2019-BLG-0335, the Einstein radius is measured, but the values of $\pie$ are not constrained.  
For the KMT-2018-BLG-0321 and KMT-2018-BLG-0885, none of the extra observables are measured, and 
only the lower limits of $\thetae$ are constrained. Due to the incompleteness of the observables, 
we estimated the physical lens parameters by conducting Bayesian analyses based on the available 
observables of the individual events.

The Bayesian analysis for each event was carried out by first generating a large number ($10^7$) 
of artificial microlensing events from a Monte Carlo simulation with the use of a prior Galactic
model. The Galactic model defines the positions, velocities, and masses of astronomical objects in
the Galaxy, and we adopted the \citet{Jung2021} model. For each simulated event, we computed the
lens observables of the Einstein time scale, $t_{{\rm E},i}=D_{{\rm L},i}\theta_{{\rm E},i}/v_{\perp,i}$, 
Einstein radius, $\theta_{{\rm E},i}=(\kappa M_i \pi_{{\rm rel},i})^{1/2}$, and microlens parallax, 
$\pi_{{\rm E},i}=\pi_{{\rm rel},i}/\theta_{{\rm E},i}$. Here, $v_\perp$ denotes the transverse 
lens-source speed. We then constructed a Bayesian posterior distributions of the lens mass and 
distance by imposing a weight to each simulated event of $w_i = \exp(-\chi^2_i/2)$, where 
$\chi^2_i=(O_i-O)^2/[\sigma(O)]^2$ and $[O, \sigma(O)]$ denote the measured value of the 
observable and its uncertainty, respectively. In the case of the event for which only the lower 
limit of $\thetae$ is constrained, we set $w_i = 0$ for events with $\thetae < \theta_{{\rm E,min}}$.

The posterior distributions for the mass of the lens companion and distance to the lens system are
presented Figures~\ref{fig:seven} and \ref{fig:eight}, respectively. In each distribution, we mark three 
curves, in which the blue and red curves represent the contributions by the disk and bulge lens populations, 
respectively, and the black curve is sum of the two contributions. The two vertical lines in the mass 
posteriors represent the boundaries between planetary, BD, and stellar lens populations. We set the 
boundary between planets and BDs as $12~M_{\rm J}$ ($M_2\sim 0.012~M_\odot$) and that between BDs 
and stars as $0.08~M_\odot$.

% Table 8 ------------------------------------------------
\begin{table}[t]
%\small
%\centering
\caption{Probabilities of the lens population and location\label{table:eight}}
\begin{tabular*}{\columnwidth}{@{\extracolsep{\fill}}l|rrr|rrr}
\hline\hline
\multicolumn{1}{c|}{Events}                &
\multicolumn{1}{c}{$P_{\rm BD}$}          &
\multicolumn{1}{c}{$P_{\rm planet}$}      &
\multicolumn{1}{c|}{$P_{\rm star}$}        &
\multicolumn{1}{c}{$P_{\rm disk}$}        &
\multicolumn{1}{c}{$P_{\rm bulge}$}       \\
\hline
KMT-2018-BLG-0321        &  59  &  3  & 38  &  75  & 25 \\ 
KMT-2018-BLG-0885        &  68  &  25 &  7  &  29  & 71 \\ 
KMT-2019-BLG-0297        &  66  &  34 &  0  &  94  &  6 \\ 
KMT-2019-BLG-0335        &  66  &  33 & 10  &  22  & 78 \\ 
%KMT-2019-BLG-3301        &  87  &  10 &  3  &  74  & 26 \\ 
\hline
\end{tabular*}
%\tablefoot{ ${\rm HJD}^\prime = {\rm HJD}- 2450000$.  }
\end{table}
% --------------------------------------------------------

In Table~\ref{table:seven}, we summarize the estimated physical lens parameters, including $M_1$, $M_2$, 
$D_{\rm L}$, and $a_\perp$, where $a_\perp=s\dl \thetae$ denotes the projected separation between the 
binary lens components. We take the median values of the posterior distributions as representative values, 
and the uncertainties are estimated as the 16\% and 84\% of the distributions. In Table~\ref{table:eight}, 
we list the probabilities for the lens companion to be in the planetary ($P_{\rm planet}$), BD 
($P_{\rm BD}$), and stellar ($P_{\rm star}$) mass regimes.  In all cases of the events, the median 
values of $M_2$ lie in the BD mass regime, and the probabilities for the lens companion to be in the 
BD mass regime are high.  For the events KMT-2018-BLG-0885, KMT-2019-BLG-0297, and KMT-2019-BLG-0335, 
the probabilities for the lens companions to be in the planetary mass regime are $P_{\rm planet}\sim 
25\%$, 34\%, and 33\%, respectively, and thus it is difficult to completely rule out the possibility 
that the companions are giant planets.  Also listed in Table~\ref{table:eight} are the probabilities 
for the lens to be in the disk, $P_{\rm disk}$, and in the bulge, $P_{\rm bulge}$.  It turns out 
that KMT-2019-BLG-0297L is very likely to be in the disk mainly from the 2-dimensional 
gaussian constraint of the measured microlens-parallax, that is, Figure~\ref{fig:four}.

\section{Summary and conclusion}\label{sec:six}

We investigated the microlensing data acquired during the 2018, 2019, and 2020 seasons by the KMTNet
survey in order to find lensing events produced by binaries with brown-dwarf companions. For this
investigation, we conducted systematic analyses of anomalous lensing events observed during the
seasons, and picked out candidate BD binary-lens events by applying the selection criterion that
the companion-to-primary mass ratio was less than 0.1. From this procedure, we identified four
candidate events with BD companions, including KMT-2018-BLG-0321, KMT-2018-BLG-0885, KMT-2019-BLG-0297, 
and KMT-2019-BLG-0335.  No candidates were identified in the 2020 season, which was severely affected 
by the Covid-19 pandemic.

We estimated the masses of the lens companions by conducting Bayesian analyses using the
measured observables of the individual events. From this estimation, it was found that the
probabilities for the masses of the companions to be in the BD mass regime were high with 
59\%, 68\%, 66\%, and 66\% for KMT-2018-BLG-0321, KMT-2018-BLG-0885, KMT-2019-BLG-0297, and 
KMT-2019-BLG-0335, respectively.  We plan to report additional BD binary-lens events from the 
investigation of the data acquired in the 2021 and 2022 seasons.  Together with the previous 6 
BD events (OGLE-2016-BLG-0890, MOA-2017-BLG-477, OGLE-2017-BLG-0614, KMT-2018-BLG-0357, 
OGLE-2018-BLG-1489, and OGLE-2018-BLG-0360) reported in paper I plus KMT-2020-BLG-0414LB 
recently reported by \citet{Zang2021}, the KMTNet sample will be useful for a future statistical 
analysis on the properties of BDs.

% --------------------------------------------------------------
\begin{acknowledgements}
Work by C.H. was supported by the grants of National Research Foundation of Korea 
(2019R1A2C2085965). 
% KMTNet
This research has made use of the KMTNet system operated by the Korea Astronomy and Space 
Science Institute (KASI) at three host sites of CTIO in Chile, SAAO in South Africa, and 
SSO in Australia. Data transfer from the host site to KASI was supported by the Korea 
Research Environment Open NETwork (KREONET). 
% MOA
The MOA project is supported by JSPS KAKENHI
Grant Number JSPS24253004, JSPS26247023, JSPS23340064, JSPS15H00781,
JP16H06287, and JP17H02871.
%Yee
J.C.Y.  acknowledges support from NSF Grant No. AST-2108414.
%Yossi Shvartzvald
Y.S.  acknowledges support from NSF Grant No. 2020740.
% China
W.Z. and H.Y. acknowledge support by the National Science Foundation of
China (Grant No. 12133005). 
% Clement Ranc
C.R. was supported by the Research fellowship of the Alexander von Humboldt Foundation.
\end{acknowledgements}

\end{document}